\documentclass[a4paper,11pt]{article}
\usepackage{pos}

\newcommand{\gev}{~\mathrm{GeV}}
\newcommand{\tev}{~\mathrm{TeV}}

\title{Novel signatures of additional Higgs bosons at the LHC}

\author*[a]{Tim Stefaniak}

\affiliation[a]{Deutsches Elektronen-Synchrotron DESY,\\
  Notkestra{\ss}e 85, 22607 Hamburg, Germany}

\emailAdd{tim.stefaniak@desy.de}

\abstract{
We identify two types of well-motivated, so-far experimentally unexplored LHC signatures of extended Higgs sectors: First, Higgs-to-Higgs decay signatures that involve (at least) two Beyond-the-Standard Model (BSM) neutral scalar bosons, $\phi_i \to \phi_j \phi_k$. These signatures are discussed in the framework of the Two-Real-Singlet-Model (TRSM), which can furthermore exhibit cascades of Higgs-to-Higgs decays leading to multi-Higgs final states. Second, the charged Higgs boson decay to a $W$ boson and a neutral BSM Higgs boson, which can easily become the dominant decay within the Two-Higgs-Doublet-Model (2HDM) parameter space. We provide benchmark scenarios for future experimental studies of these signatures.
}

\FullConference{%
  The Eighth Annual Conference on Large Hadron Collider Physics-LHCP2020 \\
  25-30 May, 2020\\
  online}

\begin{document}
\maketitle

\section{Introduction}

Many possible solutions to the open questions in particle physics (e.g., the naturalness problem, the nature of dark matter, the generation of the observed baryon asymmetry of the universe, \dots) require modifications of the Higgs sector of the Standard Model (SM). This can have the following phenomenological effects: (\emph{i}) modifications of the observed Higgs boson's properties (couplings, decay rates, CP); (\emph{ii}) presence of additional neutral or charged Higgs bosons; (\emph{iii}) presence of other new particles (e.g.\ supersymmetric particles) interacting with the Higgs boson(s). With this the Higgs sector is an exciting place to look for new physics. However, so far, experimental data has not signified any of these new physics effects.

What does the absence of serious hints for new physics in the Higgs sector, as well as in searches for other Beyond the SM (BSM) particles (supersymmetric particles, dark matter, etc.), and the stringent limits on electric dipole moments, imply for the BSM search program at the $\mathrm{TeV}$-scale? (1) Perhaps, the BSM physics scale is at a higher energy scale $\gg\mathrm{TeV}$, warranting a new collider program at the energy frontier; or (2), perhaps, the BSM signals are too faint or too rare. In this case, they may be resolved during the high-luminosity (HL) run at the LHC, and/or with improved analysis techniques and more precise theory predictions; or (3), perhaps, we have overlooked a signal in the current data. 

Here, we address the last possibility by identifying well-motivated but so-far experimentally unexplored BSM collider signatures. In particular, we highlight two types of \emph{Higgs-to-Higgs decay signatures} that have not been searched for:
\begin{enumerate}
\item Signatures that involve two BSM neutral scalar bosons, i.e., production of a scalar boson ($\phi_i$), potentially in association with other objects ($X$), $pp \to \phi_i + (X)$, followed by its decay into two lighter scalar bosons, $\phi_i \to \phi_j \phi_k$. One of the involved scalars can be identified with the observed Higgs boson ($h_{125}$). We employ the Two-Real-Singlet-Model (TRSM) to discuss these signatures, following the work of Ref.~\cite{Robens:2019kga}.
\item The production of a charged scalar boson ($\phi^\pm$), followed by its decay into a neutral non-SM scalar boson ($\phi$) and a $W^\pm$ boson, $\phi^\pm \to W^\pm \phi$. The Two-Higgs-Doublet Model (2HDM) can exhibit this signature.
\end{enumerate}

\section{$\phi_i \to \phi_j \phi_k$ in the Two-Real-Singlet-Model (TRSM)}

In the TRSM we introduce two real scalar singlet fields $S$ and $X$, leading to the scalar potential
\begin{align}
\mathcal{V} =& \mu_\Phi^2 \Phi^\dagger \Phi  + \mu_S^2 S^2 + \mu_X^2 X^2  + \lambda_\Phi (\Phi^\dagger \Phi)^2 + \lambda_S S^4 + \lambda_X X^4 + \nonumber \\
& \lambda_{\Phi S} \Phi^\dagger \Phi S^2 + \lambda_{\Phi X} \Phi^\dagger \Phi X^2 + \lambda_{SX}  S^2 X^2,
\label{eq:TRSMpotential}
\end{align}
where $\Phi$ is the Higgs doublet already present in the SM.
Here we imposed a $\mathbb{Z}_2 \times \mathbb{Z}_2' $ symmetry which is spontaneously broken by the vacuum expectation values (vevs) of the two singlet fields. We assume both $S$ and $X$ to acquire non-zero vevs. This leads to a mixing of all three neutral Higgs fields, yielding the physical states $h_1$, $h_2$ and $h_3$ with masses $M_1 \le M_2 \le M_3$ (by definition). The model parameters can be chosen to be the three masses, the three vevs and the three mixing angles, where one mass value is set to $125\gev$ and one vev is given by $246\gev$.

This simple model has a rich phenomenology of Higgs-to-Higgs decays. First, there are several mass hierarchies possible, as the observed Higgs boson can be identified with either $h_1$, $h_2$ and $h_3$.  And second, for sufficiently large mass separations, cascade decays are possible, leading to 3- or 4-Higgs final states. In particular, we can have the following signatures:
\begin{align}
h_i &\to h_j h_k \to F_\text{SM}  F_\text{SM}, \label{eq:directdecay} \\
h_3 &\to (h_2 \to h_1 h_1)\, h_{1,2} \to  F_\text{SM}  F_\text{SM}  F_\text{SM}, \label{eq:cascade1} \\
h_3 &\to (h_2 \to h_1 h_1)\, (h_2 \to h_1 h_1) \to F_\text{SM} F_\text{SM} F_\text{SM} F_\text{SM}, \label{eq:cascade2}
\end{align}
where $F_\text{SM}$ denotes a final state from a single Higgs boson decay and consists of SM particles. In fact, in the TRSM, the Higgs boson decay rates to SM particle final states are equal to those predicted in the SM, unless additional Higgs-to-Higgs decays are present and suppress these rates.

\begin{figure}
\centering
\includegraphics[width= 0.5\textwidth]{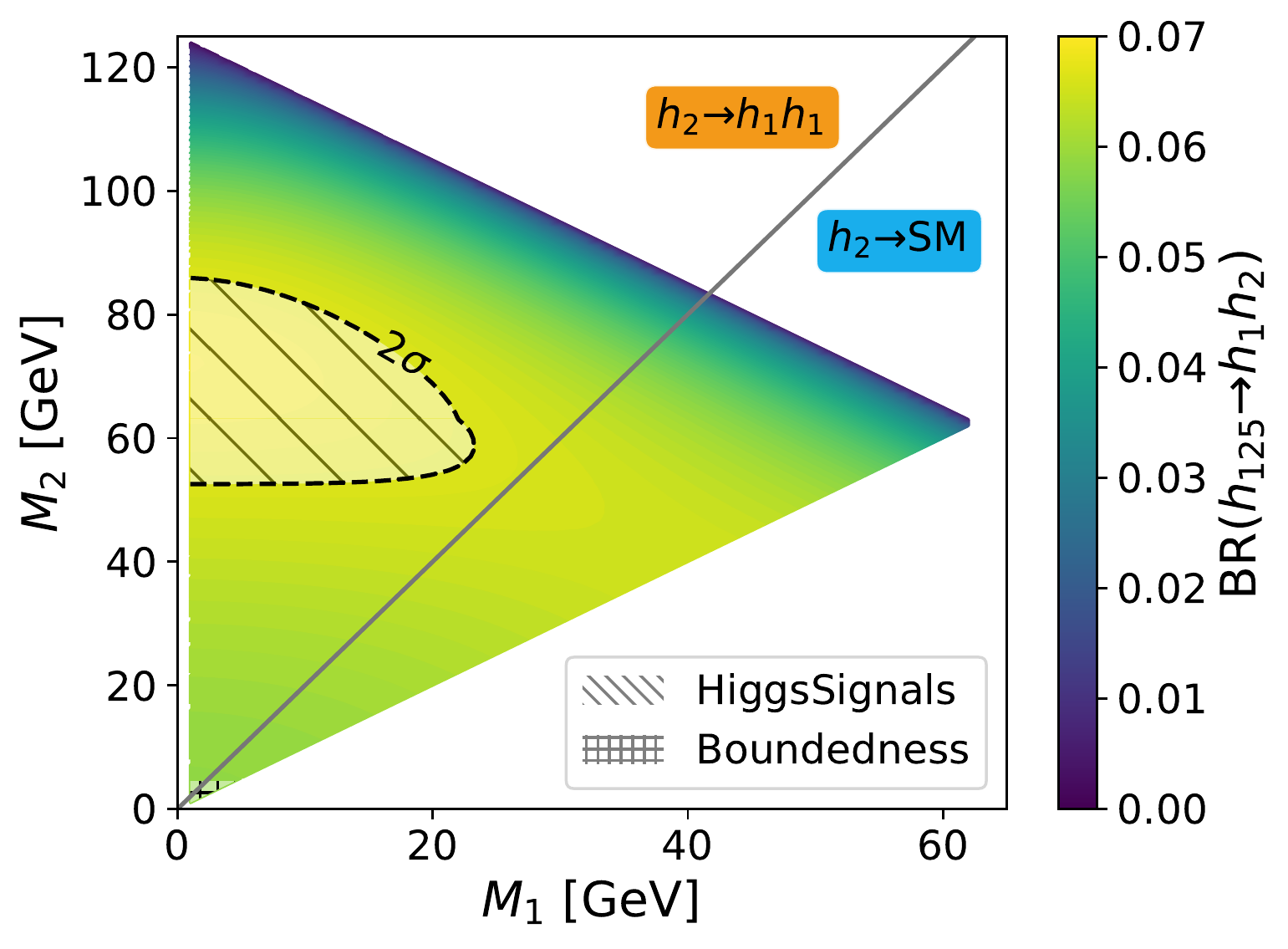}
\caption{TRSM benchmark scenario \textbf{BP1}~\cite{Robens:2019kga} for $pp \to h_3~(+X), h_3 \to h_1 h_2$, with $h_2\to F_\text{SM}$ (\emph{below the gray line}) and $h_2\to h_1 h_1$ (\emph{above the gray line}). The color scale indicates the branching  ratio $\text{BR}(h_3 \to h_1 h_2)$ and the hatched region is disfavored by Higgs rate measurements (using \texttt{HiggsSignals}~\cite{Bechtle:2013xfa}).}
\label{fig:BP1}
\end{figure}

In Ref.~\cite{Robens:2019kga} six benchmark scenarios were proposed to facilitate experimental searches for $h_i\to h_jh_k$ decays signatures. Fig.~\ref{fig:BP1} shows the ($M_1$, $M_2$) plane for one of these (\textbf{BP1}), where $h_3$ is identified with the observed Higgs state. Rates for the 4- or 6-$b$-jet final states reach up to $3~\text{pb}$ ($0.15~\text{pb}$) at the $13\tev$ LHC for $pp\to h_3$ ($pp\to V h_3$, with $V = W^\pm, Z$) production.

\section{$\phi^\pm \to W^\pm \phi$ in the Two-Higgs-Doublet-Model (2HDM)}

\begin{figure}
\centering
\includegraphics[width= 0.49\textwidth]{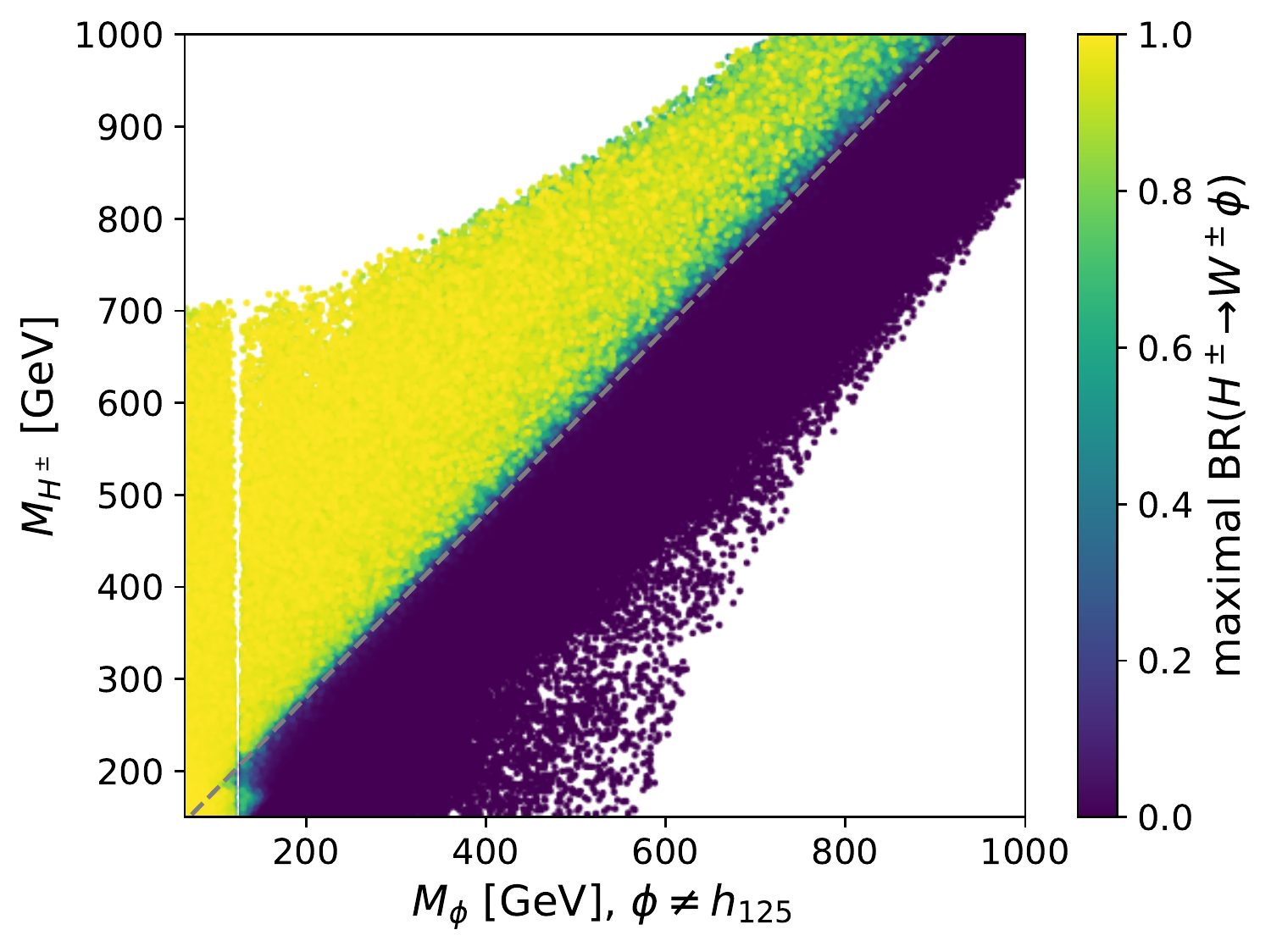} \hfill
\includegraphics[width= 0.49\textwidth]{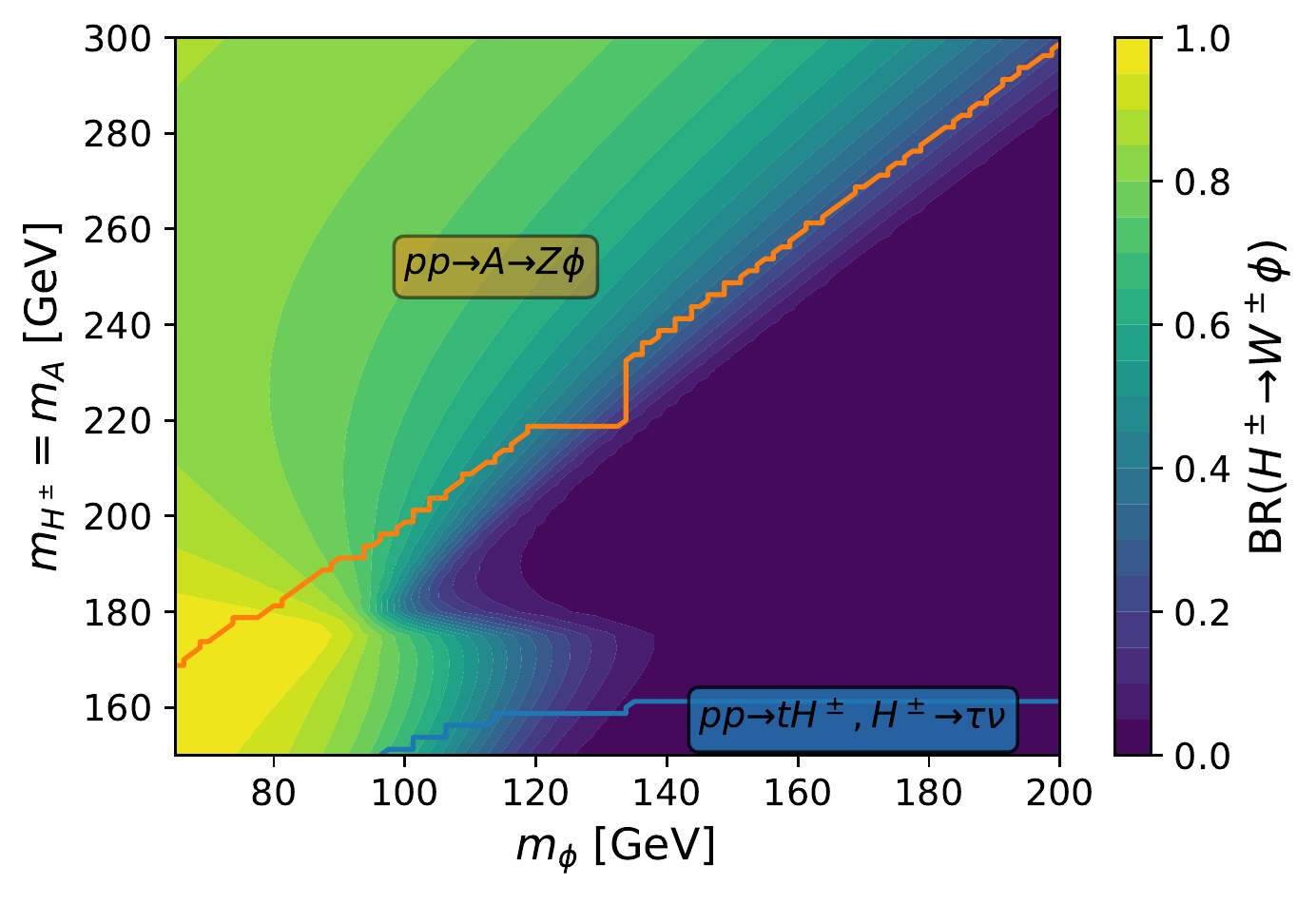}
\caption{\emph{Left:} maximal decay rate $\text{BR}(H^\pm \to W^\pm \phi)$ allowed in the 2HDM Type~1~\cite{Stefaniak2020}; \emph{right:} 2HDM benchmark scenario for experimental $H^\pm \to W^\pm \phi$ searches~\cite{Stefaniak2020}, with excluded regions from $pp\to A\to Z\phi$ and $H^\pm\to tb$ searches indicated (using \texttt{HiggsBounds}~\cite{Bechtle:2020pkv}).}
\label{fig:2HDM}
\end{figure}

Experimental searches for the charged Higgs boson, denoted $H^\pm$ in the 2HDM, mostly focus on production via $pp \to t b H^\pm$ (or, if $H^\pm$ is light enough, the top quark decay $t\to H^\pm b$) and the $H^\pm \to \tau^\pm \nu$ and $H^\pm \to t b$ decay signatures. However, as shown in Fig.~\ref{fig:2HDM} (\emph{left}), the decay $H^\pm \to W^\pm \phi$ (with a non-SM Higgs boson $\phi$), often dominates in the parameter space of the 2HDM Type~1, as the responsible coupling is maximized in the alignment limit~\cite{Arbey:2017gmh,Stefaniak:2019hvg,Stefaniak2020}. In such cases, the conventional searches for fermionic final states are insensitive.

 Fig.~\ref{fig:2HDM} (\emph{right}) shows the decay rate $\text{BR}(H^\pm \to W^\pm \phi)$ for a benchmark scenario~\cite{Stefaniak2020} defined in the $(M_\phi, M_{H^\pm} = M_A$) parameter plane with fixed parameters $\cos(\alpha - \beta) = 0~\text{or}~1$ (exact alignment), $\tan\beta =3$ and $m_{12}^2 = 500\gev^2$. Production rates for $pp \to tb H^\pm$ and $pp \to H^\pm \phi$ at the $13\tev$ LHC can reach up to $10~\text{pb}$ and $100~\text{fb}$ respectively. $\phi$ mostly decays to $b\bar{b}$ and $\tau^+\tau^-$. 
%

\section{Conclusions}

We identified two types of well-motivated, so-far experimentally unexplored \emph{Higgs-to-Higgs decay signatures}: $h_i \to h_j h_k$ decays (with possible cascades) in the TRSM, and $H^\pm \to W^\pm \phi$ in the 2HDM. We presented suitable benchmark models for future experimental studies.

\acknowledgments

We thank T.~Robens and J.~Wittbrodt for fruitful collaboration on the presented topics. This work is supported by the Deutsche Forschungsgemeinschaft (DFG, German Research
Foundation) under Germany‘s Excellence Strategy -- EXC 2121 ``Quantum
Universe'' -- 390833306.


\begin{thebibliography}{99}

\bibitem{Robens:2019kga}
T.~Robens, T.~Stefaniak and J.~Wittbrodt,
Eur. Phys. J. C \textbf{80} (2020) no.2, 151
[arXiv:1908.08554 [hep-ph]].

\bibitem{Bechtle:2013xfa}
P.~Bechtle, S.~Heinemeyer, O.~Stål, T.~Stefaniak and G.~Weiglein,
Eur. Phys. J. C \textbf{74} (2014) no.2, 2711
[arXiv:1305.1933 [hep-ph]].

\bibitem{Arbey:2017gmh}
A.~Arbey, F.~Mahmoudi, O.~St{\aa}l and T.~Stefaniak,
Eur. Phys. J. C \textbf{78} (2018) no.3, 182
[arXiv:1706.07414 [hep-ph]].

\bibitem{Stefaniak:2019hvg}
T.~Stefaniak,
[arXiv:1908.10900 [hep-ph]].

\bibitem{Stefaniak2020}
T.~Stefaniak and J.~Wittbrodt,
\emph{work in preparation.}

\bibitem{Bechtle:2020pkv}
P.~Bechtle, D.~Dercks, S.~Heinemeyer, T.~Klingl, T.~Stefaniak, G.~Weiglein and J.~Wittbrodt,
[arXiv:2006.06007 [hep-ph]].

\end{thebibliography}
\end{document}